\documentclass[aps,twocolumn,showpacs,preprintnumbers,amsmath,amssymb,superscriptaddress]{revtex4}
\usepackage{graphicx,psfrag}
\usepackage{dcolumn}
\usepackage{bm}
\usepackage[dvipsnames]{xcolor}

\begin{document}
\title{Anomalous integer quantum Hall effect in AA-stacked bilayer graphene}
\author{Ya-Fen Hsu}
\affiliation{Department of Physics
and Center for Theoretical Sciences, National Taiwan University,
Taipei 106, Taiwan}
\author{Guang-Yu Guo}
\email{gyguo@phys.ntu.edu.tw} \affiliation{Graduate Institute of Applied
Physics, National Chengchi University, Taipei 116, Taiwan}
\affiliation{Department of Physics
and Center for Theoretical Sciences, National Taiwan University,
Taipei 106, Taiwan}

\date{\today}
\begin{abstract}
Recent experiments indicate that AA-stacked bilayer graphenes (BLG) could exist.
Since the energy bands of the AA-stacked BLG are different from both the monolayer and AB-stacked
bilayer graphenes, different integer quantum Hall effect in the AA-stacked graphene is expected.
We have therefore calculated the quantized Hall conductivity $\sigma_{xy}$ and also
longitudinal conductivity $\sigma_{xx}$ of the AA-stacked BLG
within the linear response Kubo formalism.
Interestingly,
we find that the AA-stacked BLG could exhibit both conventional insulating behavior
(the $\bar{\nu}=0$ plateau) and chirality for $|\bar{\mu}|<t$, where $\bar{\nu}$ is the filling factor ($\bar{\nu}=\sigma_{xy}h/e^{2}$),
$\bar{\mu}$ is the chemical potential, and $t$ is the interlayer hopping energy,
in striking contrast to the monlayer graphene (MLG) and AB-stacked BLG.
We also find that for $|\bar{\mu}|\neq[(\sqrt{n_2}+\sqrt{n_1})/(\sqrt{n_2}-\sqrt{n_1})]t$,
where $n_1= 1, 2, 3, \cdot\cdot\cdot$, $n_2= 2, 3, 4, \cdot\cdot\cdot$ and $n_2>n_1$,
the Hall conductivity is quantized as
$\sigma_{xy}=\pm\frac{4e^2}{h}n,\mbox{ }n=0,1,2,\cdot\cdot\cdot,\mbox{ if }|\bar{\mu}|<t$ and
$\sigma_{xy}=\pm\frac{4e^2}{h}n,\mbox{ }n=1,2,3,\cdot\cdot\cdot,\mbox{ if }|\bar{\mu}|>t$.
However, if $|\bar{\mu}|=[(\sqrt{n_1}+\sqrt{n_2})/(\sqrt{n_2}-\sqrt{n_1})]t$,
the $\bar{\nu}=\pm4(n_1+n_2)n$ plateaus are absent, where $n=1,2,3,\cdot\cdot\cdot$, in comparison with
the AB-stacked BLG within the two-band approximation.
We show that in the low-disorder and high-magnetic-field regime, $\sigma_{xx}\rightarrow0$
as long as the Fermi level is not close to a Dirac point,
where $\Gamma$ denotes the Landau level broadening induced by disorder.
Furthermore, when $\sigma_{xy}$ is plotted as a function of $\bar{\mu}$, a $\bar{\nu}=0$ plateau appears across $\bar{\mu}=0$
and it would disappear if the magnetic field $B=\pi t^2/Neh\upsilon^2_F$, $N = 1, 2, 3,\cdot\cdot\cdot$.
Finally, the disappearance of the zero-Hall conductivity plateau
is always accompanied by the occurence of a $8e^2/h$-step at $\bar{\mu}=t$.
\end{abstract}
\maketitle

\section{Introduction}
Graphene exhibits many peculiar properties\cite{Castro} and has greatly intrigued physicists in recent years.
Charge carriers in the monolayer graphene (MLG) possess a linear energy dispersion (see Fig. 1a)\cite{Wallace}
and are of chiral nature\cite{Katsnelson} near each Dirac point.
The quasiparticles in the AB-stacked bilayer graphene (BLG) are also chiral.
However, unlike MLG, the energy spectra of the AB-stacked BLG are parabolic (Fig. 1b).
One of the interesting properties of graphene is quantum Hall effect (QHE).
Indeed, both theoretical\cite{Gusynin} and experimental works\cite{Novoselov1,Zhang,Novoselov2}
show that integer quantum Hall effect (IQHE) in MLG is unconventional.
The Hall conductivity in MLG is quantized as
$\sigma_{xy}=\pm\frac{4e^{2}}{h}(n+\frac{1}{2})$, where $n=0, 1, 2,\cdot\cdot\cdot$.
The factor $4$ comes from the fourfold (spin and valley) degeneracy.
Furthermore, beacuse the states are shared by electron and hole at the zeroth Landau level (LL), the shift of $1/2$ occurs.
In contrast, in the AB-stacked BLG, within the two-band parabolic approximation,
the Hall conductivity was shown to be
$\sigma_{xy}=\pm\frac{4e^{2}}{h}n$, where $n=1, 2,3,\cdot\cdot\cdot$\cite{McCann}.
The zeroth and first Landau levels are degenerate and hence the first quantum Hall plateau appears
at $4e^{2}/h$ instead of $2e^{2}/h$.
This phenomenon has been observed experimentally\cite{Novoselov3}.
QHE of the AB-stacked BLG was also studied based on a four-band Hamiltoian in Ref. \cite{Nakamura}.

Although AB stacking is predicted to be energetically favored over AA stacking in $ab$ $initio$ density functional theory
(DFT) calculations,
the energy difference of about 0.02 eV/cell is small\cite{Aoki,Andres}.
Moreover, Lauffer {\it et al.} found that scanning tunneling microscopy (STM) images of BLG
resemble that of MLG, and hence they regarded it as a consequence of the BLG confiquration being close to AA stacking\cite{Lauffer}.
Moreover, Liu {\it et al.} reported that in their high-resolution transmission electron microscope (HR-TEM) experiments
a high proportion of thermally treated samples are AA-stacked BLG\cite{Liu}.
These findings indicate the possibilty of fabrication of the AA-stacked BLG.
Since the energy bands of the AA-stacked BLG (see Fig. 1c) are different from both
the AB-stacked BLG and monolayer graphene, the quantum Hall effect in
the AA-stacked is expected to be quite different from that in the latter two systems.

We have therefore carried out a theoretical study of IQHE as well as the longitudinal
conductivity $\sigma_{xx}$ in the AA-stacked bilayer graphene using the Kubo formalism.
In this paper, we present a general analytical form of the
Hall conductivity [$\sigma_{xy}(\bar{\mu},B)$]
as a function of both chemical potential ($\bar{\mu}$) and magnetic field ($B$)
of the AA-stacked BLG. Our presentation will be divided into two parts:
i) the variation of the $\sigma_{xy}$ vs. $1/B$ curve with some fixed $\bar{\mu}$'s and
ii) the effect of the magnetic field on the $\sigma_{xy}$ vs. $\bar{\mu}$ curve.
Our main findings are as follows.
Firstly, for $|\bar{\mu}|\neq[(\sqrt{n_2}+\sqrt{n_1})/(\sqrt{n_2}-\sqrt{n_1})]t$,
where $t$ is the interlayer hopping energy, $n_1$ and $n_2$ are any integers larger than 1 and 2, respectively, and $n_1 < n_2$, 
the Hall conductivity is quantized as
\begin{align}
&\sigma_{xy}=\pm\frac{4e^2}{h}n,\mbox{ }n=0,1,2,\cdot\cdot\cdot,\mbox{ if }|\bar{\mu}|<t\notag\\
&\sigma_{xy}=\pm\frac{4e^2}{h}n,\mbox{ }n=1,2,3,\cdot\cdot\cdot,\mbox{ if $|\bar{\mu}|>t$}.
\end{align}
However, if $|\bar{\mu}|=[(\sqrt{n_2}+\sqrt{n_1})/(\sqrt{n_2}-\sqrt{n_1})]t$, the Hall conductivity is given by
\begin{equation}
\sigma_{xy}=\pm\frac{4e^2}{h}n\mbox{, excluding }\pm\frac{4e^{2}}{h}(n_1+n_2)n,\mbox{ }n=1,2,3,\cdot\cdot\cdot.\
\end{equation}
Secondly, in the low-disorder and high-magnetic-field regime [$\Gamma\rightarrow0$ and $\hbar^4\omega_c^4\gg(\bar{\mu}+\nu t)^4$],
$\sigma_{xx}\approx(8e^2/\pi h)[(\bar\mu^2+t^2)\Gamma^2/(\bar{\mu}^2-t^2)^2
+2\Gamma^2/\hbar^2\omega_c^2+5(\bar{\mu}^2+t^2)\Gamma^2/\hbar^4\omega_c^4)]\rightarrow0$ if the Fermi
level is not close to a Dirac point, where $\hbar\omega_c$ is the cyclotron energy.
That is to say, if the magnetic field is high enough, the applied electric field cannot drive any
current for $|\bar{\mu}|<t$, while the
current is perpendicular to external electric field for $|\bar{\mu}|>t$.
Thirdly, we find that the $\sigma_{xy}=0$ (the filling factor $\bar{\nu}=0$) plateau
across $\bar{\mu}=0$ would disappear when $B=\pi t^2/Neh\upsilon^2_F$, $N = 1, 2, 3,\cdot\cdot\cdot$,
and that a $8e^2/h$-step at $\bar{\mu}=t$ occurs while a zero-Hall conductivity plateau disappears.
Interestingly, unlike the monolayer and AB-stacked bilayer graphenes,
the AA-stacked bilayer graphene could display a unusual $\bar{\nu}=0$ plateau even though it contains chiral quasiparticles.
We argue that the occurrence of the $\bar{\nu}=0$ plateau is due to the shift of level anomalies by the interlayer hopping energy $t$.

\section{Theoretical model and analytical calculation}
\subsection{Model Hamiltonian and Landau levels}

\begin{figure}
\includegraphics[width=8cm]{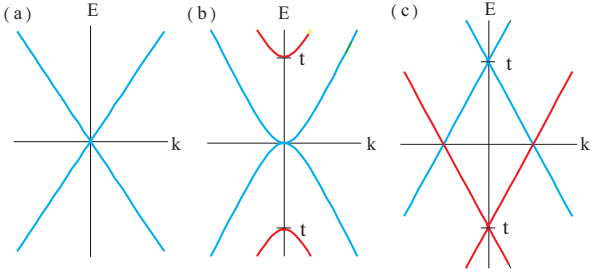}
\caption{(color online) The energy bands of (a) monolayer graphene, (b) AB-stacked bilayer graphene,
and (c)AA-stacked bilayer graphene.}
\end{figure}
We first derive an effective four-band Hamiltonian near each Dirac point for the AA-stacked bilayer graphene
in tight-binding aprroximation via ${\bf k}\cdot{\bf p}$ expansion\cite{Castro}.
We then diagonalize this Hamiltonian to obtain the energy bands of the AA-stacked BLG.
In Fig. 1, we display the energy bands of the MLG, AB-stacked, and AA-stacked BLG together.
We find that the energy bands of the AA-stacked BLG are just two copies of the MLG band structure
shifted up and down by $t$, respectively. 
Hence, $E=\pm t$ are the Dirac points for the AA-stacked BLG.
When a magnetic field is applied, we should replace the momentum operator ${\bf p}$ with ${\bf p}+e{\bf A}/c$,
where the external magnetic field ${\bf B}=\nabla \times{\bf A}$ and -e is the charge of an electron.
The magnetic field is applied along the positive $z$-axis (i.e. out of plane) and hence
the vector potential can be written as ${\bf A} =(-By,0,0)$ in the Landau gauge.
Therefore, the effective four-band Hamiltonian in the presence of the magnetic field is given by
\begin{align}
H_{\pm}&= \left(\begin{array}[c]{cc}
   \upsilon_{F}(\sigma_{x}\pi_{x}\pm\sigma_{y}\pi_{y}) & -tI\\
   -tI & \upsilon_{F}(\sigma_{x}\pi_{x}\pm\sigma_{y}\pi_{y})
   \end{array} \right)\notag\\
&:\left\{\begin{array}{ll}
\pi_{x}=-i\hbar\partial_{x}-eBy/c, \\[2pt]
\pi_{y}=-i\hbar\partial_{y}. \\[2pt]
\end{array}\right.
\end{align}
Here $\pm$ label the two valleys of the band structure at $K$ and $K'$, respectively.
$\upsilon_{F}$ denotes the Fermi velocity.
In the Laudau gauge, we can substitute the eigenfunction $\psi=e^{ikx}\phi(y)$
of the Hamiltonian into the schr\"{o}dinger equation $H\psi=E\psi$.
Here $\phi(y)$ can be written as $(\phi_{1}(y),\phi_{2}(y))^{T}$, where $\phi_{1}$ and $\phi_{2}$ are two-component column vectors.
Then, we make the transformations: $\sigma^{\pm}=\sigma_{x}{\pm}i\sigma_{y}$, $\xi=y/\ell_{B}-\ell_{B}k$ and $O^{\mp}=(\xi\pm\partial_{\xi})/\sqrt{2}$
, where the magnetic length $\ell_{B}=\sqrt{{\hbar}c/|eB|}$.
Finally, for the $K$ valley, the schr\"{o}dinger equation reads
\begin{equation*}
\left(\begin{array}[c]{cc}
   \frac{-\hbar\upsilon_{F}}{\sqrt{2}\ell_{B}}(O^{-}\sigma^{+}+O^{+}\sigma^{-}) & -tI\\
   -tI & \frac{-\hbar\upsilon_{F}}{\sqrt{2}\ell_{B}}(O^{-}\sigma^{+}+O^{+}\sigma^{-})
   \end{array} \right)
\left(\begin{array}[c]{cc}
        \phi_{1}\\\phi_{2}
        \end{array}\right)
\end{equation*}
\begin{equation}
=E\left(\begin{array}[c]{cc}
        \phi_{1}\\\phi_{2}
        \end{array}\right).
\end{equation}

Since $O^{\mp}$ satisfy the commutation relation: $[O^{-}, O^{+}]$$=1$, $O^{\mp}$ are
the annihilation and creation operators of one-dimensional (1-D)
simple harmonic oscillator (SHO), respectively.
Similarly, $\sigma^{\pm}$ are the raising and lowering operators of pesudospin angular momentum.
Obviously, the eigenstates of $O^{-}\sigma^{+}+O^{+}\sigma^{-}$ are $(|N-1\rangle,\pm|N\rangle)^{T}$ if $N\geq1$
and $(0,|0\rangle)^T$ if $N=0$,
where $|N\rangle$ are the eigenstates of the 1-D SHO.
All non-zero vectors are eigenvectors of $-tI$.
Therefore, we can infer that the eigenvalues of Eq. (4) are
\begin{equation}
E^{\mu\nu}_{N}=-\mu\sqrt{N}\hbar\omega_{c}-{\nu}t
\end{equation}
with $\omega_{c}=\sqrt{2}\upsilon_{F}/{\ell_{B}}$ and the eigenstates of Eq. (4) are
\begin{equation}
\left\{\begin{array}{ll}
|0,+,\nu\rangle=\frac{1}{\sqrt{2}}\left(
\begin{array}[c]{cc}
0\\
|0\rangle\\
0\\
\nu|0\rangle
\end{array}\right), & \mbox{if }N=0, \\[2pt]
|N,\mu,\nu\rangle=\frac{1}{2}\left(
\begin{array}[c]{cc}
|N-1\rangle\\
\mu|N\rangle\\
\nu|N-1\rangle\\
\mu\nu|N\rangle
\end{array}\right), & \mbox{if }N\geq1,
\end{array} \right.
\end{equation}
Here the indice $\mu=\pm$ and $\nu=\pm$.
Clearly, the LLs of the AA-stacked BLG are just two copies of the LLs of the MLG
shifted up and down by $t$, respectively.

\subsection{Linear response calculation}
The conductivity can be calculated using the Kubo formula within the linear response theory\cite{Bruus}.
The Kubo formula for the DC-conductivity is given by
\begin{equation}
\sigma_{ij}=\lim_{\Omega\rightarrow0}\frac{\mbox{Im}\Pi^{R}_{ij}(\Omega+i0)}{\hbar\Omega}.
\end{equation}
Here the retarded current-current correlation $\Pi^{R}_{ij}$ in the Matsubara form reads\cite{Nakamura}
\begin{align}
&\Pi^{R}_{ij}(i\nu_{m})=-\frac{4e^{2}}{2\pi\ell^{2}_{B}\beta\hbar}\times\\
&\sum^{\infty}_{n=-\infty}\sum^{\infty}_{k,\ell=0}\sum_{\scriptstyle \mu,\rho \atop =\scriptstyle \pm}\sum_{\scriptstyle \nu,\sigma \atop =\scriptstyle \pm}
\frac{{\langle}k,\mu,\nu|v_{i}|\ell,\rho,\sigma\rangle\langle\ell,\rho,\sigma|v_{j}|k,\mu,\nu\rangle}
{(i\tilde{\omega}_{n}-\tilde{E}_{k}^{\mu\nu})(i\tilde{\omega}_{n}+i\nu_{m}-\tilde{E}_{\ell}^{\rho\sigma})}, \notag
\end{align}
where the factor $4$ is due to the fourfold (spin and valley) degeneracy,
the velocity operator $v_{i}=[x_{i}\mbox{ }H_{+}]/i\hbar$, and $\tilde{E}^{\mu\nu}_{k}={E}^{\mu\nu}_{k}/\hbar$.
$\omega_{n}$ and $\nu_{m}$ are Matsubara frequencies of fermion and boson, respectively.
When the chemical potential and disorder scattering are considered,
Matsubara frequency of fermion has to be corrected as $i\tilde{\omega}_{n}=i\omega_{n}+\bar{\mu}/\hbar+i\mbox{sgn}({\omega}_{n})\Gamma/\hbar$.
$\Gamma$ is the Landau level broadening due to the presence of disorder.
Substituting the eigenstates of $H_{+}$ into Eq. (8), we obtain
\begin{equation}
\Pi^{R}_{xy}(i\nu_{m})=-\frac{ie^{2}v_{F}^{2}}{2\pi\ell^{2}_{B}\beta\hbar}
(2\sum_{\scriptstyle \mu,\nu \atop =\scriptstyle \pm}\chi^{\mu\nu,+\nu}_{1,0}
+\sum_{\ell\geq1}\sum_{\scriptstyle \mu,\rho \atop =\scriptstyle \pm}\sum_{\nu=\pm}
\chi^{\mu\nu,\rho\nu}_{\ell+1,\ell}),
\end{equation}
where $\chi^{\mu\nu,\rho\sigma}_{k,\ell}$ is defined as
\begin{align}
\chi^{\mu\nu,\rho\sigma}_{k,\ell}(i\nu_{m})&=\sum^{n=\infty}_{n=-\infty}[\frac{1}{(i\tilde{\omega}_{n}-\tilde{E}_{k}^{\mu\nu})
(i\tilde\omega+i\nu_{m}-\tilde{E}_{\ell}^{\rho\nu})}\notag\\
&-(i\nu_{m}\rightarrow-i\nu_{m})].
\end{align}

In the DC and clean limit, we let $\Omega+i0\rightarrow0$ and set $\Gamma=0$. 
Then, after evaluating the Matsubara sums\cite{Bruus}, we find that
\begin{equation}
\frac{1}{\beta\hbar}\chi^{\mu\nu,\rho\sigma}_{k,\ell}|_{i\nu_{m}=\Omega+i0\rightarrow0}\thickapprox
\frac{-2(\Omega+i0)[f(E^{\mu\nu}_{k})-f(E^{\rho\sigma}_{\ell})]}{(\tilde{E}^{\mu\nu}_{k}-\tilde{E}^{\rho\sigma}_{\ell})^{2}},
\end{equation}
where $f(E)$ is the Fermi-Dirac distribution.
Furthermore, we define $\tilde{f}(E)=f(E)-1/2$.
Then, using Eqs. (7), (9) and (11) and considering the variation of direction of conductivity with the signs of magnet field and carrier charge,
we can derive that
\begin{equation}
\sigma_{xy}=-\frac{4e^{2}}{h}\mbox{sgn}(eB)\left[\sum_{\nu=\pm}\tilde{f}(E_{0}^{+\nu})+
\sum_{\ell\geq1}^{\infty}\sum_{\scriptstyle \mu,\nu \atop =\scriptstyle \pm}
\tilde{f}(E^{\mu\nu}_{\ell})\right]
\end{equation}

At zero temperature, $f(E)=1$ and $f(E)=0$ for the occupied and unoccupied LLs, respectively.
The LLs located at $E\leq\bar{\mu}$ are ocuppied and the others are empty.
That is to say, $\tilde{f}(E)=1/2$ for $E\leq\bar{\mu}$, while $\tilde{f}(E)=-1/2$ for $E>\bar{\mu}$.
Therefore, we only need to calculate the number of LLs between $-\bar{\mu}$ and $\bar{\mu}$ to
determine the magnitude of the Hall conductivity.
Moreover, for $-|\bar{\mu}|<E<|\bar{\mu}|$, the number of the up-shifted LLs ($\nu=-$) is equal to that of the down-shifted LLs ($\nu=+$).
Hence, we find that the zero-temperature Hall conductivity is given by
\begin{align}
\sigma_{xy}=& -\frac{4e^{2}}{h}\mbox{sgn$(\bar{\mu})$sgn$(eB)$}\left[\theta(|\bar{\mu}|+t)\theta(|\bar{\mu}|-t)\right.\notag\\
&+\sum_{\ell\geq1}^{\infty}\sum_{\mu=\pm}\theta(|\bar{\mu}|-E_{\ell}^{\mu-})
\theta(|\bar{\mu}|+E_{\ell}^{\mu-})\left.\right].
\end{align}
Eq. (13) indicates that the Hall conductivity would simply change its sign as $\bar{\mu}\rightarrow-\bar{\mu}$ and
that the Hall conductivity is equal to $4e^2/h$ times the number of up-shifted LLs between $-|\bar{\mu}|$ and $|\bar{\mu}|$.
Eq. (13) can also be written in the form
\begin{align}
&\sigma_{xy}=-\frac{4e^{2}}{h}\mbox{sgn}(\bar{\mu})\mbox{sgn}(eB)\times\notag\\
&\left\{\theta(|\bar{\mu}|-t-\sqrt{2{\hbar}v_{F}^{2}|eB|/c})\left[\frac{c(|\bar{\mu}|-t)^{2}}{2{\hbar}v_{F}^{2}|eB|}\right]\right.\notag\\
&+\theta(\sqrt{2{\hbar}v_{F}^{2}|eB|/c}-t+|\bar{\mu}|)\left[\frac{c(|\bar{\mu}|-t)^{2}}{2{\hbar}v_{F}^{2}|eB|}\right] \\
&+\left[\frac{c(|\bar{\mu}|+t)^{2}}{2{\hbar}v_{F}^{2}|eB|}\right]-\left[\frac{c(|\bar{\mu}|-t)^{2}}{2{\hbar}v_{F}^{2}|eB|}\right]\notag\\
&+\left.\theta(|\bar{\mu}|+t)\theta(|\bar{\mu}|-t)\right\}\notag.
\end{align}
Here $[x]$ means the integer part of $x$.
The last term is the contribution of level anomalies $(N=0)$\cite{Gusynin}.

We also calculate the longitudinal conductivity ($\sigma_{xx}$) via the Kubo formula.
The longitudinal conductivity have to be evaluated under the effect of disorder scattering ($\Gamma\neq0$).
Using Cauchy's integral theorem as in Refs. \cite{Mahan,Iwasaki}, we can obtain
\begin{align}
&\sigma_{xx}=\frac{e^2}{{\pi}h}\omega_{c}^{2}\left\{\int^\infty_{-\infty}dE\left(-\frac{{\partial}f}{{\partial}E}\right)\times\right.\notag\\
&\left.\left[\sum_{\scriptstyle \mu,\rho \atop =\scriptstyle \pm}
\sum_{\nu=\pm}\sum_{\ell\geq1}^{\infty}\mbox{Im}\tilde{g}_{\ell+1}^{\mu\nu}(E)\mbox{Im}\tilde{g}_{\ell}^{\rho\nu}(E)
+2\sum_{\scriptstyle \mu,\nu \atop =\scriptstyle \pm}\mbox{Im}\tilde{g}_{1}^{\mu\nu}(E)\mbox{Im}\tilde{g}_{0}^{+\nu}(E)\right]\right\},
\end{align}
where $\tilde{g}_{\ell}^{\rho}=1/(E/\hbar-\tilde{E}_{\ell}^{\rho\nu}+i\Gamma/\hbar)$.
Applying the techniques of partial-fraction decomposition similar to that used in Ref. \cite{Endo} and after some cumbersome algebra,
we finally find that the zero-temperature longitudinal conductivity can be written in terms of the digamma function ($\varphi_{0}$)\cite{Lebedev},
\begin{align}
&\sigma_{xx}= \frac{4e^2}{{\pi}h}\sum_{\nu=\pm}\left\{i\left[\varphi_{0}\left(-\frac{(\bar{\mu}+{\nu}t)^2-\Gamma^2}
{\hbar^2\omega_c^2}+1-i\frac{2\Gamma(\bar{\mu}+{\nu}t)}{\hbar^2\omega_c^2}\right)\right.\right.\notag\\
&-\left.\varphi_{0}\left(-\frac{(\bar{\mu}+{\nu}t)^2-\Gamma^2}
{\hbar^2\omega_c^2}+1+i\frac{2\Gamma(\bar{\mu}+{\nu}t)}{\hbar^2\omega_c^2}\right)\right]\notag\\
&\times\frac{\left[2(\bar{\mu}+{\nu}t)^3\Gamma+2(\bar{\mu}+{\nu}t)\Gamma^3\right]}{p}\notag\\
&+\frac{\left[\hbar^6\omega_c^6+\hbar^4\omega_c^4(\bar{\mu}+{\nu}t)^2-12\hbar^2\omega_c^2(\bar{\mu}+{\nu}t)^4+8(\bar{\mu}+{\nu}t)^6\right]\Gamma^2}
{q}\notag\\
&+\frac{\left[\hbar^4\omega_{c}^4+4\hbar^2\omega_{c}^2(\bar{\mu}+{\nu}t)^2+16(\bar{\mu}+{\nu}t)^4\right]\Gamma^4}{q}\notag\\
&+\frac{8(\bar{\mu}+{\nu}t)^2\Gamma^6}{q}\notag\\
&+\frac{\left[\hbar^2\omega_c^2(\bar{\mu}+{\nu}t)^2+\hbar^4\omega_c^4\right]\Gamma^2+\hbar^2\omega_c^2\Gamma^4}
{\left[(\bar{\mu}+\hbar\omega_c+{\nu}t)^2+\Gamma^2\right]\left[(\bar{\mu}-\hbar\omega_c+{\nu}t)^2+\Gamma^2\right]}\notag\\
&\times\left.\frac{1}{(\bar{\mu}+{\nu}t)^2+\Gamma^2}\right\}.
\end{align}
Here,
\begin{align}
&p=\hbar^4\omega_{c}^4+16\Gamma^2(\bar{\mu}+{\nu}t)^2,\notag\\
&q=\left\{[(\bar{\mu}+{\nu}t)^2-\hbar^2\omega_{c}^2-\Gamma^2]^2+4\Gamma^2(\bar{\mu}+{\nu}t)^2\right\}\times p.
\end{align}
Unlike Ref. \onlinecite{Gusynin2}, we do not adopt low-magnetic field approximation here
and therefore Eq. (16) is general and suitable for all values of the applied magnetic field.
However, when we adopt the low-disorder and high-magnetic-field limit,
we first consider $\Gamma$ as being small and keep the terms in the order of $\Gamma^2$.
Then, we let $(\bar{\mu}+\nu t)/(\hbar^4\omega_c^4)\rightarrow0$, Eq. (16) could be simplified as
\begin{equation}
\sigma_{xx}\approx\frac{4e^2}{\pi h}\left[\frac{(\bar{\mu}^2+t^2)\Gamma^2}{(\bar{\mu}^2-t^2)^2}
+2\frac{\Gamma^2}{\hbar^2\omega_c^2}+5\frac{\bar{\mu}^2+t^2}{\hbar^2\omega_c^2}\frac{\Gamma^2}{\hbar^2\omega_c^2}\right].
\end{equation}
The first term is independent of $B$.
The second and third terms are proportional to $1/B$ and $1/B^2$, respectively.

\begin{figure}
\includegraphics[width=7.0cm]{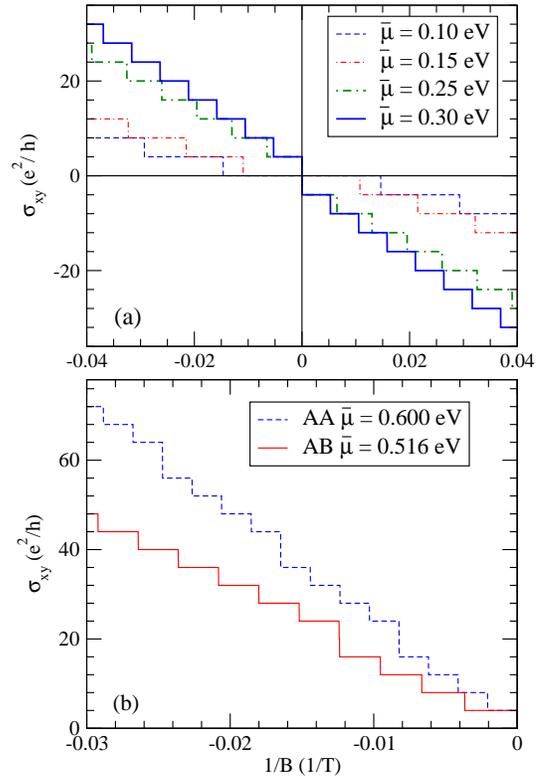}
\caption{(color online) (a) The quantized Hall conductivity $\sigma_{xy}$ of the AA-stacked bilayer graphene
as a function of $1/B$ for several values of chemical potential $\bar{\mu}$.
(b) The quantized Hall conductivity $\sigma_{xy}$ of both AA-stacked and AB-stacked bilayer graphenes
as a function of $1/B$. 
The interlayer hopping energy $t$ used is 0.2 eV for the AA-stacking and 0.4 eV for the AB-stacking,
and $\upsilon_{F}=1.0\times10^6$ m/s. The Hall conductivity $\sigma_{xy}$ for the
AB-stacking was obtained by using Eq. 15 from Ref. \onlinecite{Nakamura}.}
\end{figure}

\section{Dependence of conductivity on magnetic field and chemical potential}
Since both the magnetic field and chemical potential could be tuned experimentally,
we display the calculated conductivity as a function of $1/B$ and $\bar{\mu}$ in this Sec.
Here we use the interlayer hopping energy $t=0.2$ eV for the AA-stacking and $t=0.4$ eV for the AB-stacking
as determined by the $ab$ $initio$ DFT calculations within the local density approximation (LDA)\cite{Yeh}. 
Moreover, because the quantized values of the Hall conductivity is independent of the presence of disorder scattering\cite{Hajdu},
we show only the Hall conductivity in the clean limit (i.e., $\Gamma=0$) and analyze the Hall plateaus qualitatively.

\subsection{Conductivities vs inverse of magnetic field} 
In Fig. 2, the Hall conductivity is plotted as a function of $1/B$
and we should discuss the effect of chemical potential on the $\sigma_{xy}$ vs. $1/B$ curve.
From Fig. 2(a), we see that $\sigma_{xy}=\pm\frac{4e^{2}}{h}n$, with $n=0,1,2,\cdot\cdot\cdot$, for $|\bar{\mu}|<t$,
and $n=1,2,3,\cdot\cdot\cdot$, for $|\bar{\mu}|>t$,
excluding $|\bar{\mu}|=[(\sqrt{n_2}+\sqrt{n_1})/(\sqrt{n_2}-\sqrt{n_1})]t$,
where $n_1=1,2,3,\cdot\cdot\cdot$ and $n_2=2,3,4,\cdot\cdot\cdot$.
It is clear from either Eq. (13) or Eq. (14) that $\sigma_{xy}(-\bar{\mu})=-\sigma_{xy}(\bar{\mu})$,
and hence we did not show any curves for $\bar{\mu}<0$ in Fig. 2(a).
Interestingly, in constrast to the MLG and AB-stacked BLG,
the AA-stacked BLG displays the pronounced $\bar{\nu}= 0$ plateau for $|\bar{\mu}|<t$,
where the filling factor $\bar{\nu}=\sigma_{xy}h/e^{2}$.
The MLG and AB-stacked BLG lack the  $\bar{\nu}=0$ plateau because their level anomalies are located at $E=0$.
The level anomaly of the MLG is the zeroth Landau level
while those of the AB-stacked BLG are the zeroth and first Landau levels.
The occurrence of level anomalies is a remarkable manifestation of the unique property
of chiral quasiparticles\cite{McCann,Yoshioka}.
In the  AA-stacked BLG, similarly, level anomalies also exist in the LL spectrum.
However, for the AA-stacked BLG, the level anomalies are shifted up and down by $t$, respectively.
Thus, these level anomalies are unique in the sense that
they are always located at the Dirac points regardless of the magnitude of $B$,
in contrast with the other Landau levels\cite{Gusynin,McCann,Nakamura}.
It is worth mentioning that the Dirac points often exhibit interesting electronic properties,
such as electron-hole puddle formation\cite{Hwang, Martin}, and Andreev reflection type transitions.\cite{Yafen}
Untill now, some transport properties at these Dirac points remain to be understood\cite{Geim}.
The level anomaly is one of the interesting properties of the Dirac points and
recently the nature of its electronic states (being metallic or insulating) in
the high magnetic field-low temperature regime is hotly debated\cite{LZhang}.

\begin{figure}
\includegraphics[width=7.0cm]{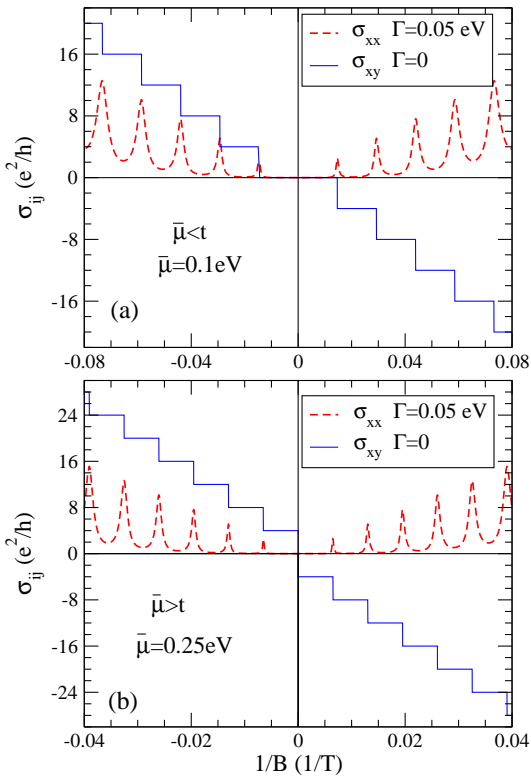}
\caption{(color online) The longitudinal ($\sigma_{xx}$) and transverse ($\sigma_{xy}$) conductivities
of the AA-stacked bilayer graphene as a fuction of $1/B$,
(a) for $|\bar{\mu}|<t$, and
(b) for $|\bar{\mu}|>t$.
The rest parameters are the same as in Fig. 2.}
\end{figure}

Displayed in Fig. 2(b) are the Hall plateaus of the AA-stacked BLG for $|\bar{\mu}|=[(\sqrt{n_2}+\sqrt{n_1})/(\sqrt{n_2}-\sqrt{n_1})]t$.
We find that in addition to the $\bar{\nu}=0$ plateau, other differences between the AA-stacked and AB-stacked BLGs exist.
In particular, when $|\bar{\mu}|=[(\sqrt{n_2}+\sqrt{n_1})/(\sqrt{n_2}-\sqrt{n_1})]t$, some $8e^2/h$-steps appear at $1/B\neq0$.
In other words, in comparison with the other cases of $|\bar{\mu}|>t$ ($|\bar{\mu}|\neq[(\sqrt{n_1}+\sqrt{n_2})/(\sqrt{n_2}-\sqrt{n_1})]$),
some plateaus would be missing for $|\bar{\mu}|=[(\sqrt{n_2}+\sqrt{n_1})/(\sqrt{n_2}-\sqrt{n_1})]t$.
Furthermore, these $8e^2/h$-steps appear periodically.
Taking $\bar{\mu}=3t$ (i.e. $n_2=4, n_1=1$), for example, between any two $8e^2/h$-steps,
the curve passes through three $4e^2/h$-steps and four plateaus.
Only the Hall plateaus $\sigma_{xy}=\frac{4e^2}{h}n$,
$n=\pm1,\pm2,\pm3,\pm4,\pm6,\pm7,\pm8,\pm9,\pm11,\pm12,\cdot\cdot\cdot$, appear.
In other words, the Hall plateaus $n=\pm5,\pm10,\pm15,\cdot\cdot\cdot$ are absent here.
When $|\bar{\mu}|>t$, the quantum Hall effect of the AB-stacked BLG must be studied based on a four-band Hamiltonian\cite{Nakamura}.
Based on the four-band model\cite{Nakamura}, for $|\bar{\mu}|>t$, the AB-stacked BLG can also exhibit a $8e^2/h$-step,
as shown in Fig. 2(b). Although the $8e^2/h$-step is not specific to the AA-stacked BLG,
the periodic appearance of the $8e^2/h$-steps has never been seen in the AB-stacked BLG
and hence is a unique characteristic of the AA-stacked BLG.

The longitudinal and transverse conductivities as a fuction of $1/B$ are plotted together in Fig. 3.
It is seen from Fig. 3 that the longitudinal conductivity goes to a local minima at the position of the Hall plateaus
and reaches a local maxima as the steps appear except the step at $1/B=0$.
The unique $\bar{\nu}=0$ Hall plateau of the AA-stacked bilayer graphene for $|\bar{\mu}|<t$ is especially interesting.
From Fig. 3(a), we find that $\sigma_{xx}$ falls to zero as the $\bar{\nu}=0$ Hall plateau emerges.
As stated previously, for $|\bar{\mu}|>t$, the AA-stacked BLG lacks the $\bar{\nu}=0$ Hall plateau.
The first Hall plateau occurs at $\bar{\nu}=4$ or $\bar{\nu}=-4$.
This encourages us to investigate the difference between the longitudinal conductivities at the $\bar{\nu}=0$ Hall plateau for $|\bar{\mu}|<t$
and at the $\bar{\nu}=\pm4$ Hall plateaus for $|\bar{\mu}|>t$.
Fig. 3(b) shows that for $|\bar{\mu}|>t$, the $\sigma_{xx}$ goes to zero as the first Hall plateau occurs.
This implies that at the high magnetic field, the external electric field cannot drive any current
for $|\bar{\mu}|<t$ while the current is perpendicular to the external electric field for $|\bar{\mu}|>t$.
Here $\bar{\mu}$ and $t$ are in the order of 0.1 eV while the order of magnitude of $\Gamma$ [$\mathcal{O}(\Gamma)$] is 0.01 eV.
$\bar{\mu}+\nu t$ are about one order of magnitude higher than $\Gamma$ ($\bar{\mu}+\nu t\sim10\Gamma$).
Therefore, the condition of low disorder is satisfied.
If $\hbar^4\omega_c^4\gg(\bar{\mu}+\nu t)$, Eq. (18) can be applied here.
This needs $\hbar\omega_c$ to be larger than 1.78 $(\bar{\mu}+\nu t)$
and $B$ is estimated to be at least a few times larger than 10 T.
In the low-disorder and high-magnetic-field regime, we roughly estimate from Eq. (18)
\begin{equation}
\sigma_{xx}\sim\frac{4e^2}{h}\left[\frac{2}{\pi}\frac{(\bar{\mu}^2+t^2)\Gamma^2}{(\bar{\mu}^2-t^2)^2}+\mathcal{O}(0.001)\right].
\end{equation}
When $\mathcal{O}(|\bar{\mu}|-t)\ge1\Gamma$, $\sigma_{xx}\sim0.1(4e^2/h)\rightarrow0$.

Interestingly, the results shown in Fig. 2 could be explained in terms of Fig. 4.
Let us account for Fig. 2(a) first.
It is clear from Fig. 4 that for $|\bar{\mu}|<t$, level anomalies are outside the range of $-|\bar{\mu}|\sim|\bar{\mu}|$.
Conversely, for $|\bar{\mu}|>t$, level anomalies are inside the range of $-|\bar{\mu}|\sim|\bar{\mu}|$.
In the high magnetic field regime, for $|\bar{\mu}|<t$, no Landau level exists between $-|\bar{\mu}|$ and $|\bar{\mu}|$
and hence the AA-stacked BLG displays the conventional insulating behaviour (a $\bar{\nu}=0$ Hall plateau)
even though it possesses chirality.
Such behaviour of the AA-stacked BLG is in stark contrast to the MLG and AB-stacked BLG.
However, for $|\bar{\mu}|>t$, level anomalies are located between $-|\bar{\mu}|$ and $|\bar{\mu}|$ even in the high magnetic field regime
and hence a $\bar{\nu}=0$ plateau cannot emerge.

The Hall plateaus displayed in Fig. 2(b) can be explained as follows.
A up-shifted LL of $E_{k}^{-\mu,-}$ and a down-shifted LL of $E_{k}^{\mu,+}$ are partners
because $E_{k}^{\mu,+}=-E_{k}^{-\mu,-}$.
As the up-shifted LL of $E_{k}^{-\mu,-}$ goes through the $|\bar{\mu}|$-level,
the down-shifted LL of $E_{k}^{\mu,+}$ passes through the $-|\bar{\mu}|$-level.
They always enter the region between $|\bar{\mu}|$ and $-|\bar{\mu}|$ (i.e. the shaded region) together
and hence each contribute $4e^2/h$ to the Hall conductivity.
Therefore, the Hall conductivity is equal to $4e^2/h$ times the number of either up-shifted
or down-shifted LLs between $-|\bar{\mu}|$ and $|\bar{\mu}|$.
Therefore, in order to form a $8e^2/h$-step, either two up-shifted or down-shifted LLs
must enter or leave the shaded region together.
Hence we only need to focus on either up-shifted or down-shifted Dirac cones and discuss the movement of
the LLs located in this cone to explain the origin of $8e^2/h$-steps.
Let us consider the up-shifted Dirac cone.
As the magnetic field decreases gradually, the up-shifted LLs would go close to its level anomaly, $E=t$.
Therefore, we can infer that for $|\bar{\mu}|<t$,
the LLs above the $|\bar{\mu}|$-level (called the upper LLs) go far away from the $|\bar{\mu}|$-level,
while the LLs below the $-|\bar{\mu}|$-level (called the lower LLs) move toward the $-|\bar{\mu}|$-level,
as shown in Fig. 4(a).
The lower LLs would enter the shaded region one by one but the upper LLs can never get into the shaded region.
However, for $|\bar{\mu}|>t$, both the upper and lower LLs can go close to the shaded region [see Fig. 4(b)],
i.e., two up-shifted LLs may enter the shaded region together.
Therefore, the $8e^2/h$-steps can only appear when $|\bar{\mu}|>t$.

\begin{figure}
\includegraphics[width=8.0cm]{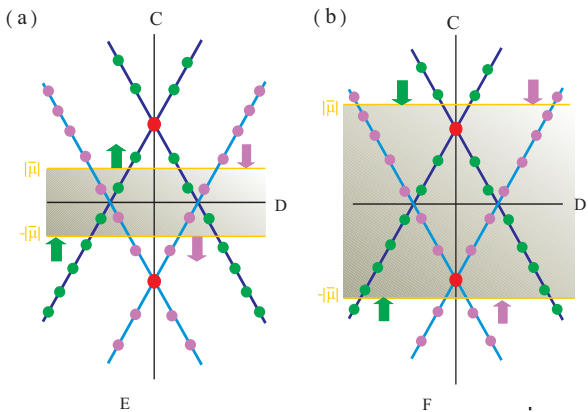}
\caption{(color) The Landau level spectrum of the AA-stacked bilayer graphene
(a) with $|\bar{\mu}|<t$ and (b) with $|\bar{\mu}|>t$, respectively.
Here circles mark the positions of Landau levels.
Red, green, and purple circles represent the locations of level anomalies as well as
other up-shifted and down-shifted Landau levels, respectively.
Mazarine and blue lines denote the up-shifted and down-shifted energy bands, respectively.
Shaded is the region between $-\bar{\mu}$ and $|\bar{\mu}|$.}
\end{figure}

For brevity, we use indices $(k,\mu,\nu)$ to denote the Landau level of $E_{k}^{\mu,\nu}$ below.
Let us label the two LLs which enter the shaded region together as the $(k_1,-,-)$ and $(k_2,+,-)$ LLs.
Then, $k_1/k_2=(|\bar{\mu}|-t)^2/(|\bar{\mu}|+t)^2$.
$(n_1,n_2)$ satisfies this condition and $n_1/n_2$ is an irreducible fraction.
Then, $(k_1,k_2)=(pn_1,pn_2)$, where $p=1,2,3\cdot\cdot\cdot$, is a set of solutions of $k_1/k_2=(|\bar{\mu}|-t)^2/(|\bar{\mu}|+t)^2$.
Between the entries of $[(p-1)n_1,(p-1)n_2]$ and $[pn_1,pn_2]$ LLs,
$(n_1+n_2-2)$ LLs get into the shaded region sequentially as $1/B$ decreases.
Hence, $(n_1+n_2-2)$ $4e^2/h$-steps occur between any two $8e^2/h$-steps.
The Hall conductivity is quantized as $\sigma_{xy}=\pm\frac{4e^{2}}{h}n$
with the exception of $\pm\frac{4e^{2}}{h}(n_1+n_2)n$, where $n=1, 2,3,\cdot\cdot\cdot$, as shown in Fig. 2(b).
Unlike the AB-stacked BLG and the other cases of $|\bar{\mu}|>t$ of the AA-stacked BLG, 
the Hall conductivity for $|\bar{\mu}|=[(\sqrt{n_2}+\sqrt{n_1})/(\sqrt{n_2}-\sqrt{n_1})]t$ lacks the $\bar{\nu}=\pm4(n_1+n_2)n$ plateaus.
Furthermore, it is clear from Fig. 2(b) that when $|\bar{\mu}|>t$, the AB-stacked BLG can also exhibit a $8e^2/h$-step
but the appearance of the $8e^2/h$-steps is not periodical.
The main findings here are summarized in Table I.


\begin{table}
\caption{The characteristics of integer quantum Hall effect in monolayer (ML), AB-stacked bilayer (AB)
and AA-stacked
bilayer (AA) graphenes as well as conventional two-dimensional semiconductor structures (2D).}
\begin{ruledtabular}
\begin{tabular}{ccccc}
                 & ML & AB & AA & 2D \\ \hline
  plateau steps ($\frac{e^2}{h}$)  & 4 & 4, 8\footnotemark[1] & 4, 8\footnotemark[2] & 2  \\
 $\bar{\nu} = 0$ plateau  & no  & no & yes\footnotemark[3] & yes \\
\end{tabular}
\footnotetext[1]{The 8$\frac{e^2}{h}$ plateau step only occurs (aperiodically) in the four band model (Ref. \cite{Nakamura}).}
\footnotetext[2]{The 8$\frac{e^2}{h}$ plateau step appears periodically for $|\bar{\mu}|>t$ only.}
\footnotetext[3]{The $\bar{\nu} = 0$ plateau occurs only if $|\bar{\mu}|<t$.}
\end{ruledtabular}
\label{tableD}
\end{table}

\subsection{Hall conductivity vs chemical potential}

\begin{figure}
\includegraphics[width=8.0cm]{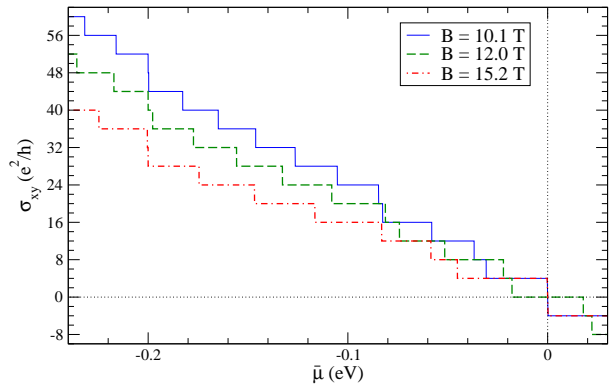}
\caption{(color online) The quantized Hall conductivity $\sigma_{xy}$ of the AA-stacked bilayer graphene
as a function of chemical potential $\bar{\mu}$ for several values of magnetic field $B$.
The rest parameters are the same as in Fig. 2.}
\end{figure}

Fig. 5 is a plot of $\sigma_{xy}$ versus $\bar{\mu}$, showing how the Hall plateaus are influenced by the magnetic field.
It is clear from Fig. 4 that unlike the MLG and AB-stacked BLG,
a $\bar{\nu}=0$ plateau centered at $\bar{\mu}=0$ appears for $B=12$ T in the AA-stacked BLG.
However, when the condition of $\sqrt{2N}\hbar\upsilon_{F}/\ell_{B}=t$ is reached by tuning the magnetic field,
the $(N,+,-)$ and $(N,-,+)$ LLs would be located at $E=0$. As a result, the $\bar{\nu}=0$ plateau disappears and a $8e^2/h$-step at $\bar{\mu}=0$ forms, like the AB-stacked BLG.
In other words, the absence of the $\bar{\nu}=0$ plateau needs the magnetic field
$B=\pi t^2/Neh\upsilon^2_F$, where $N=1,2\cdot\cdot\cdot$.
In Fig. 5, $B=10.1$ T and $B=15.2$ T satisfy this condition with $N=3$ and $N=2$, respectively.
Thus, these curves lack the  $\bar{\nu}=0$ plateau.

In addition, we note that a $8e^2/h$-step occurs at $\bar{\nu}=t$ for $B=10.1$ T and $B=15.2$ T.
For $B=12$ T, all the LLs are nondegenerate and hence all the steps are of the height of $4e^2/h$.
However, if the $(N,-,+)$ and $(o, +,-)$ LLs are degenerate at $E=t$, a $8e^2/h$-step appears at $\bar{\mu}=t$.
This level degeneracy happens as the magnetic field $B=4\pi t^2/Neh\upsilon^2_F$, where $N=1,2,3,\cdot\cdot\cdot$.
$B=10.1$ T and $B=15.2$ T fit the condition with $N=3$ and $N=2$, respectively,
and thus a $8e^2/h$-step appears at $\bar{\mu}=t$.
We also find that the disappearance of a zero-Hall conductivity plateau is always accompanied by the occurence of a $8e^2/h$-step at $\bar{\mu}=t$,
because if $B=\pi t^2/Neh\upsilon^2_F$, $B$ would satisfy $B=4\pi t^2/N'eh\upsilon^2_F$ with $N'=4N$.
Interestingly, here we find that the structure of the Hall plateaus of the AA-stacked BLG would be significantly affected by the
applied magnetic field, which is quite different from the MLG and the AB-stacked BLG.

\section{summary}
In conclusion, we have calculated both the quantized Hall conductivity and longitudinal conductivity
of the AA-stacked bilayer graphene within linear response theory by using Kubo formula.
We find that the dependence of the Hall plateau of the AA-stacked BLG on the magnetic field
is distinctly different from both the MLG and AB-stacked BLG as well as the conventional quantun
Hall materials.
In particular, the AA-stacked bilayer graphene could possess the unique $\bar{\nu}=0$ plateau,
in contrast to other graphene materials such as monolayer and AB-stacked bilayer graphene.
The shift of level anomalies due to interlayer hopping energy is attributed to be the origin of the $\bar{\nu}=0$ plateau.
Nonetheless, the $\bar{\nu}=0$ plateau across $\bar{\mu}=0$
would disappear if magnetic field $B=\pi t^2/Neh\upsilon^2_F$.
In addition, we find that the disappearance of a zero-Hall conductivity plateau
is always accompanied by the occurence of a $8e^2/h$-step at $\bar{\mu}=t$.
Furthermore, when $|\bar{\mu}|=[(\sqrt{n_1}+\sqrt{n_2})/(\sqrt{n_2}-\sqrt{n_1})]t$,
the AA-stacked BLG lacks the $\bar{\nu}=\pm4(n_1+n_2)n$ plateaus, which exist in the AB-stacked BLG.
We also find that when $\Gamma\rightarrow0$ and $\hbar^4\omega_c^4\gg(\bar{\mu}+\nu t)^4$, $\sigma_{xx}\rightarrow0$ if
the Fermi level is not a few $\Gamma$s above and below a Dirac point.
This implies that at the high magnetic field, the external electric field cannot drive any current
for $\bar{\mu}<t$ while
the current would be perpendicular to the external electric field for $\bar{\mu}>t$.
We hope that our predicted interesting characteristics of quantum Hall effect in the AA-stacked
BLG, which are not seen in both the MLG and BLG as well as the conventional quantum Hall
materials, would stimulate experimental effort on fabricating the AA-stacked BLG and also
on measurement of its transport property in the near future.

\section*{ACKNOWLEGEMENTS}
The authors thank Hung-Yu Yeh, Tsung-Wei Chen and Ming-Che Chang for valuable discussions.
The authors also acknowledge financial supports from National Science Council and NCTS of Taiwan.


\begin{references}
\bibitem{Castro}A. H. Castro Neto, N. M. R. Peres, K. S. Novoselov, and A. K. Geim, Rev. Mod. Phys. $\bf 81$, 109 (2009).

\bibitem{Wallace}P. R. Wallace, Phys. Rev. $\bf 71$, 622 (1947)

\bibitem{Katsnelson} M. I. Katsnelson, K. S. Novoselov, And A. K. Geim, Nat. Phys. $\bf 2$, 620 (2006).

\bibitem{Gusynin} V.P. Gusynin, and S.G. Sharapov, Phys. Rev. Lett. $\bf 95$, 146801 (2005).

\bibitem{Novoselov1}K. S. Novoselov, A. K. Geim, S. V. Morozov, D. Jiang, M. I. Katsnelson, I. V. Grigorieva, S. V. Dubonos and A. A. Firsov, Nature (London)
$\bf 438$, 197 (2005).

\bibitem{Zhang} Y. Zhang, Y. W. Tan, H. L. Stormer, and P. Kim, Nature (London) ${\bf 438}$, 201 (2005).

\bibitem{Novoselov2} K.S. Novoselov,Z. Jiang, Y. Zhang, S. V. Morozov, H. L. Stormer, U. Zeitler, J. C. Maan, G. S. Boebinger, P. Kim, and A.K. Geim,
Science $\bf 315$, 1379 (2007).

\bibitem{McCann} E. McCann and V. I. Fal'ko, Phys. Rev. Lett. ${\bf 96}$, 086805 (2006).

\bibitem{Novoselov3} K. S. Novoselov1, E. McCann, S. V. Morozov, V. I. Fal'ko, M. I. Katsnelson, U. Zeitler, D. Jiang, F. Schedin and A. K. Geim,
Nat. Phys. $\bf 2$, 177 (2006).

\bibitem{Nakamura} M. Nakamura, L. Hirasawa, and K.-I. Imura, Phys. Rev. B $\bf 78$, 033403 (2008).

%
%
%
%
%
\bibitem{Aoki} M. Aoki and H. Amawashi, Solid State Commun. $\bf 142$, 123(2007).

\bibitem{Andres} P. L. de Andres, R. Ram\'{1}rez, and J. A. Verg\'{e}s, Phys. Rev. B $\bf 77$, 045403 (2008).

\bibitem{Lauffer} P. Lauffer, K. V. Emtsev, R. Graupner, Th. Seyller, and L. Ley, Phys. Rev. B ${\bf 77}$, 155426 (2006).

\bibitem{Liu} Z. Liu, K. Suenagam, H. Suzuura, P. J. F. Harris, and S. Iijima, Phys. Rev. Lett. ${\bf 102}$, 015501 (2009).
%
%
%
%
%
%

\bibitem{Bruus} H. Bruus, and K. Flensberg, Many-Body Quantum Theory in Condensed Matter Physics (Oxford, New York, 2007), Chaps. 6 and 11.
%

\bibitem{Mahan} G. D. Mahan, Many-Particle Physics (Plenum Press, New York, 2000), Chap. 4.

\bibitem{Iwasaki} M. Iwasaki, H. Ohnishi and T. Fukutome, J. Phys. G: Nucl. Part. Phys. $\bf 35$, 035003 (2008).

\bibitem{Endo} A. Endo, N. Hatano2, H. Nakamura, and R. Shirasaki, J. Phys. : Condens. Matter $\bf 21$, 345803 (2009).

\bibitem{Lebedev} N. N. Lebedev, Special Functions and Their Applications (Dover, New York, 1972), Chap. 1.

\bibitem{Gusynin2} V. P. Gusynin, and S. G. Sharapov, Phy. Rev. B $\bf 71$, 125124 (1972).

\bibitem{Yeh} H. Y. Yeh (private communication ).

\bibitem{Hajdu} M. Jan$\beta$en, O. Veihweger, U. Fastenrath, and J. Hajdu, Introduction to the Theory of the Integer Quantum Hall
Effect (VCH, Weinheim, 1994).

\bibitem{Yoshioka} D.Yoshioka, The Quantum Hall effect (Spring-Verlag, Berlin Heidelberg, 2002), Chap. 7.

\bibitem{Hwang} E. H. Hwang, S. Adam, and S. Das Saima, Phys. Rev. Lett. $\bf 98$, 186806 (2007).

\bibitem{Martin} J. Martin, N. Akerman, G. Ulbricht, T. Lohmann, J. H. Smet, K. Von Klitzing, and A. Yacoby, Nat. Phys. $\bf 4$, 144 (2008).

\bibitem{Yafen} Y. F. Hsu, and G. Y. Guo, Phys. Rev. B $\bf 81$, 045412 (2010).

\bibitem{Geim} A. K. Geim, Science $\bf 324$, 1530 (2004).

\bibitem{LZhang} L. Zhang, Y. Zhang, M. Khodas, T. Valla, and I. A. Zaliznyak, arxiv:1003.2738 (2010).

\end{references}
\end{document}